\documentclass[a4paper]{article}
\usepackage{amsmath}
\usepackage{amssymb}
\usepackage{graphicx}
\usepackage{setspace}
\usepackage{hyperref}
\usepackage{xcolor}
\usepackage[a4paper]{geometry}
\usepackage{changepage}
\usepackage{caption}
\usepackage{adjustbox}
\usepackage{comment}
\usepackage{float}
\usepackage{braket}
\usepackage{graphicx,booktabs,array}
\usepackage{lineno}

\usepackage{subfig}
\usepackage{xcolor}
\usepackage[normalem]{ulem}

\newcommand{\authors}[2]{%
\begin{center}
{\large #1 \\ \textit{#2}}
\end{center}}

\renewcommand{\title}[1]{%
\begin{center}
{\Large\textbf{#1}}
\end{center}}

\begin{document}

\title{\Large{Dynamic Generation of Photonic Spatial Quantum States with an All-Fiber Platform
}}

\begin{center}
\authors{\textbf{{A. Alarcón$^{1,2,*}$}, J. Argillander$^{1,2}$, D. Spegel-Lexne$^1$, and G.B. Xavier$^1$}}
\footnotesize{{$^1$\textit{Institutionen f\"{o}r Systemteknik, Link\"{o}pings Universitet, SE-581 83 Link\"{o}ping, Sweden\\} }}

\footnotesize{{$^2$\textit{The authors contributed equally to this work\\} }}
\footnotesize{{$^*$\textcolor{blue}{alvaro.alarcon@liu.se} }}

\end{center}

\begin{abstract}
    Photonic spatial quantum states are a subject of great interest for applications in quantum communication. One important challenge has been how to dynamically generate these states using only fiber-optical components. Here we propose and experimentally demonstrate an all-fiber system that can dynamically switch between any general transverse spatial qubit state based on linearly polarized modes. Our platform is based on a fast optical switch based on a Sagnac interferometer combined with a photonic lantern and few-mode optical fibers. We show switching times between spatial modes on the order of 5 ns and demonstrate the applicability of our scheme for quantum technologies by demonstrating a measurement-device-independent (MDI) quantum random number generator based on our platform. We run the generator continuously over 15 hours acquiring over 13.46 Gbits of random numbers, of which we ensure that at least 60.52\% are private following the MDI protocol. Our results show the use of photonic lanterns in order to dynamically create spatial modes using only fiber components, which due to its robustness and integration capabilities have important consequences for photonic classical and quantum information processing. 
\end{abstract}

\section{Introduction}

In quantum technologies individual quantum and entangled states are employed for information processing, providing significant advantages in areas such as computing and communication security \textbf{[1, 2]}. The latter has motivated the integration of quantum communication technology with current classical communication systems in order to share existing telecom infrastructure \textbf{[3, 4]}. 
In order to make this integration possible, quantum communications systems must be compatible with the telecommunication optical spectrum, the employed components and optical fibers. Single-mode fibers (SMF) have long been, and continue to be, the preferred platform for quantum communication experiments \textbf{[5, 6]}. However, the growing demand for data traffic has forced us to rethink the existing technological infrastructure \textbf{[7]}. For instance, spatial division multiplexing (SDM) technologies make it possible to use the transverse spatial properties of a light beam to multiplex information and increase data transport capacity \textbf{[8]}, as well as offering advantages for other fields such as in quantum information \textbf{[4]}.

These recent technological advances have motivated various studies that have allowed a quantum system to be encoded in terms of its transverse optical spatial modes \textbf{[9-14]}. One type of SDM fiber that has gained considerable attention recently are few-mode fibers (FMFs) \textbf{[15]}, specially due to their ability to support orbital angular momentum modes \textbf{[11, 13, 16, 17]}. The FMF core is slightly larger cross-sectionally than the core of an SMF, allowing it to carry more than one transverse spatial mode. The main approach is to prepare $N$ spatial modes and propagate an $N$-dimensional qudit represented by a spatial superposition of Gaussian modes through the FMF. However, a common problem in quantum communication with transverse spatial modes is that it usually relies on bulk optical components such as q-plates \textbf{[18]}, spatial light modulators \textbf{[19]}, or cylindrical lens mode converters \textbf{[20]}, employed to manipulate the spatial states, which all suffer from slow response times, thus limiting the performance in communication applications and hindering integration with fiber-optic systems. 

In order to start handling these issues, several fiber-based techniques have been implemented to create spatial superpositions of light, such as the use of long-period gratings (LPGs) \textbf{[21 - 23]} or mode-selective couplers \textbf{[24, 25]}. Even though the generation efficiency of spatial modes in these devices is higher than that of bulk components, they are still usually based on manual tuning. One promising technique is based on creating parallel path states of light (with a beamsplitter for instance), which are then converted into superpositions of linearly polarized (LP) modes with a photonic lantern (PL) and then launched onto a few-mode fiber \textbf{[13, 16]}. The PL has been used to increase the capacity of telecommunications channels through the multiplexing of multiple transverse spatial channels in the same fiber \textbf{[7, 8]}. Our commercial PL (Phoenix Photonics) has an FMF at the output and 3 independent single-mode fibers at the input. We can assign each of these Gaussian input modes to one of the three lowest order LP modes in the FMF. The internal structure is made of an adiabatic taper that provides a slow transition from the input single-mode fibers to the FMF in such a way that it transforms the single-mode input to the proper LP mode at the output. The spatial modes in the FMF will be given by the interaction of the supermodes in the internal structure of the PL, and the mapping is produced by a matching process between the effective indices of the tapered region and the incoming spatial modes \textbf{[16, 26]}. This device has also shown remarkable capability to integrate with other SDM technologies and fiber optic communication systems \textbf{[7, 26]}, which makes it a great candidate for generating spatial modes. In this paper we show that a Sagnac interferometer acting as a tunable beamsplitter combined with a photonic lantern can be used to dynamically, and at high-speed, generate any two-dimensional spatial state of the form $\alpha|0\rangle + \beta e^{i\phi}|1\rangle$, with $\alpha, \beta \in [0, 1]$ where $|0\rangle$ and $|1\rangle$ correspond to two orthogonal LP modes in the FMF. It is worth noting in particular that we can also generate the specific superpositions $|\text{OAM}_{\pm}\rangle = 1/\sqrt{2}(|\text{LP}_\text{11a}\rangle + e^{\pm i\pi/2}|\text{LP}_\text{11b}\rangle)$, which are the orbital angular momentum modes (OAM) of index $\pm1$, currently under intense investigation in optical and quantum communication systems \textbf{[3]}. Our system can switch between two orthogonal spatial states with a speed on the range of 5 ns, limited by the driving electronics. We demonstrate the applicability of our system by dynamically preparing different quantum states as required in a measurement-device-independent (MDI) quantum random number generator (QRNG) \textbf{[27]}. We demonstrate the stability of our system by continuously measuring over 15 hours with an average random number generation rate of $205.82\pm 19.95$ kbps of data which, after randomness extraction, passes the widely adopted NIST 800-22 test suite \textbf{[28]}. Another recent work was able to demonstrate the feasibility of generating random numbers from superpositions of transverse spatial states over ring-core fibers (RCFs) \textbf{[29]}, with a proposal to upgrade it to a measurement device-independent case based on optical switches on an integrated photonic circuit. Our results go beyond that by already demonstrating the MDI protocol thanks to our dynamic spatial state preparation method, and also without the need of a long fiber to generate temporal delays among the spatial states. We have used only off-the-shelf components, which shows the feasibility of employing FMF technology in applications in quantum information. Our results show a reliable and fast technique to prepare transverse spatial modes of light in optical fibers, with potential applications in classical and quantum information fields. 

\section{Setup}

The experimental setup to prepare the transverse spatial states of light is shown in Fig.\textbf{[1]} a). The weak coherent source (WCS) consists of a continuous DFB semiconductor laser ($\lambda$= 1546.92 nm), followed by a fiber pigtailed lithium niobate intensity modulator (IM) and then a variable optical attenuator in order to produce weak coherent states (not shown for the sake of clarity). The weak coherent states then go to an optical circulator (C) before entering one of the input ports of a 50:50 bidirectional fiber coupler (FC). The Sagnac loop (SL) is constructed by connecting the outputs of the FC together with a lithium niobate pigtailed telecom phase modulator (PM), used to give a phase shift $\phi_R$, and a 50 m long fiber optic spool as a delay line. The phase modulator is used to change the relative phase between the two counter propagating directions inside the loop, and the delay line is needed to ensure sufficient time separation to apply adequate phase modulation to adjust for different output probabilities of the loop. This is achieved by correctly choosing the relative delay between the IM within the SPS and the PM within the loop. A field-programmable gate array (FPGA) applies an electrical signal to the PM to ensure that only the wavepacket propagating in the clockwise direction is subjected to a phase shift. Two manual polarization controllers are used to align the polarization of the wave packets traveling clockwise and counterclockwise in order to maximize the interference when they are recombined in the bidirectional fiber coupler. 

\begin{figure}[ht]
\centering
 
    \includegraphics[width=1\textwidth]{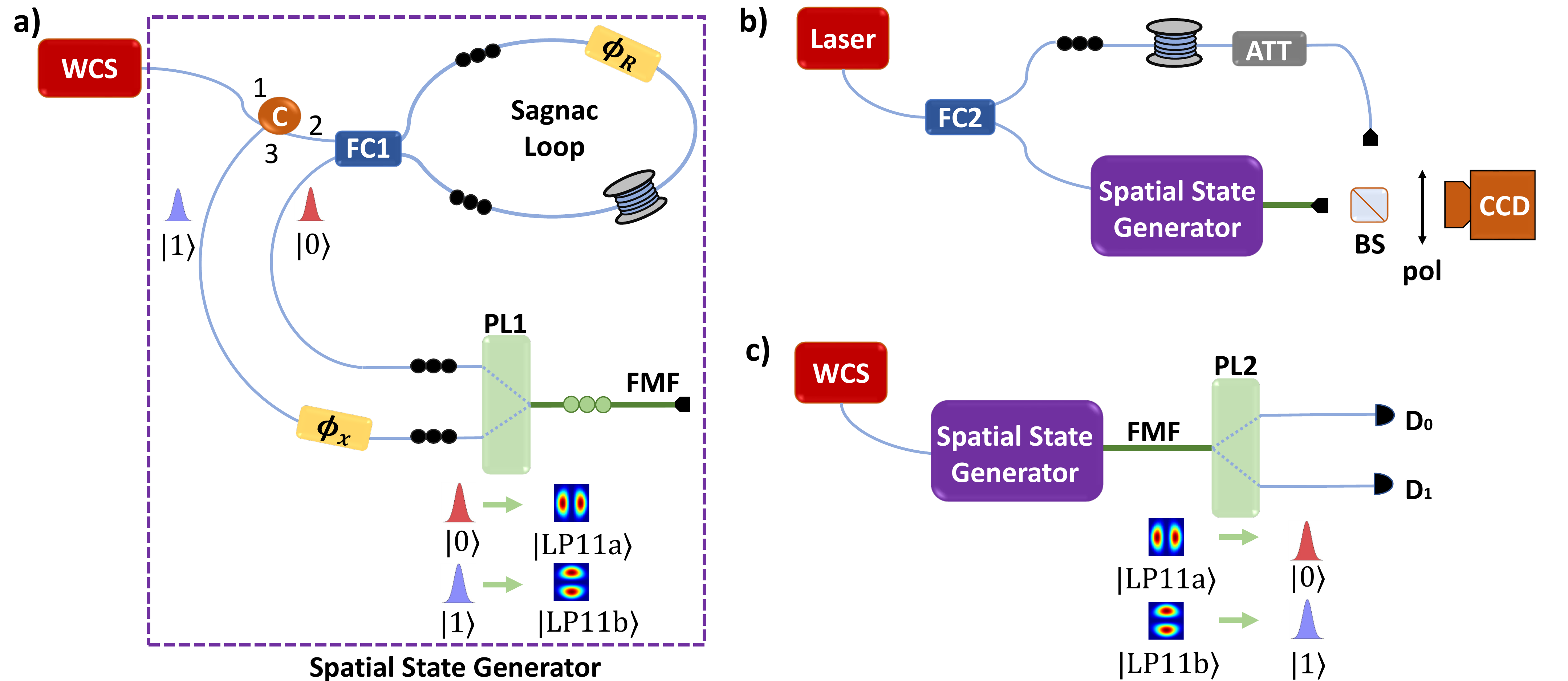}
    \caption{ Experimental setup. a) Our transverse spatial state generator (delimited in the dashed box). We employ a Sagnac loop acting as a tunable beamsplitter to create superpositions of the $|0\rangle$ and $|1\rangle$ path states. By combining this with a relative phase shift $\phi_x$, we can then create any general superposition of the form $\alpha|0\rangle + \beta e^{i\phi_x}|1\rangle$, with $\alpha$ and $\beta$ determined by the internal phase setting $\phi_R$ in the Sagnac loop. The path state is then mapped to the general superposition of linearly polarized modes $\alpha|\text{LP}_\text{11a}\rangle + \beta e^{i\phi_x}|\text{LP}_\text{11b}\rangle$ in the FMF by the photonic lantern. b) We employ an infrared CCD camera to characterize the amplitude and the interferogram of the generated spatial states. c) We demonstrate a measurement-device-independent quantum random number generator by using our state generator to dynamically switch between $|\text{LP}_\text{11a}\rangle$, $|\text{LP}_\text{11b}\rangle$ and its superposition. The projection is done on the path basis by converting from the LP states back to the path states, and detecting the outputs with single-photon detectors. ATT: Optical attenuator; BS: Beamsplitter; C: Optical circulator; CCD: Charged coupled device camera; D: Single-photon detector or amplified p-i-n photodiode; FC: Bidirectional fiber coupler; FMF: Few-mode fiber; PL: Photonic lantern; Pol: Optical polarizer; WCS: Weak coherent source
 \label{fig:figure1}} 
\end{figure}

We create the well known general path state ${|\psi\rangle} = \alpha|{0}\rangle + i \beta|{1}\rangle $, where $|{0}\rangle$ and $|{1}\rangle$ denote the path states at the output of the SL. $\alpha=\frac{1}{2}\left(1-e^{i\phi_R} \right)$ and $\beta=\frac{1}{2} \left(1+e^{i\phi_R} \right)$ are the probability amplitudes of the photon to be routed through $|{0}\rangle$ and $|{1}\rangle$ (after the circulator) respectively. Thus the SL operates as a tunable beamsplitter controlled by $\phi_R$ \textbf{[35]}. One of the paths is connected to another pigtailed lithium niobate phase modulator ($\phi_x$) to create the superposition ${|\psi\rangle} = \alpha|{0}\rangle + e^{i\phi_x} \beta|{1}\rangle $. Both paths are then connected to a mode-selective photonic lantern (PL1) whose main function is to take $N$ input single-mode fibers, and map each one of them into $N$ corresponding LP modes in a FMF \textbf{[26]}.  

In our case PL1 makes the following mapping: $|{0}\rangle \rightarrow |\text{LP}_\text{11a}\rangle$ and $|{1}\rangle \rightarrow |\text{LP}_\text{11b}\rangle$.  Two additional manual polarization controllers are placed on each path $|{0}\rangle$ and $|{1}\rangle$ to optimize the modal excitation at the output of PL1, which consists of an FMF capable of supporting three spatial modes of propagation: $|\text{LP}_\text{01}\rangle$, $|\text{LP}_\text{11a}\rangle,$ and $|\text{LP}_\text{11b}\rangle$. The mapping performed by the photonic lantern creates the state $|\psi\rangle= \alpha|\text{LP}_\text{11a}\rangle + e^{i \phi_x} \beta|\text{LP}_\text{11b}\rangle$ in the FMF. By acting on the phase modulators $\phi_x$ and $\phi_R$ it is possible to create any state on the surface of the Bloch sphere.

\section{Creating arbitrary spatial superpositions}

In order to demonstrate that we can generate any arbitrary photonic state $\psi$, we employ the setup in Fig. Fig.\textbf{[1]} b). We remove the attenuation from the WCS such that it is now working at standard power levels. Furthermore the pulse width is adjusted to be 100 ns to take into account the slower response time of the InGaAs CCD camera used to measure the intensity and phase profiles of the generated states. 
The generated light pulses following the IM are split in two paths with a 50:50 fiber coupler, with the lower path connected to our dynamic spatial state generator producing the state $|\psi\rangle$, which is then collimated with a 10x microscope objective and launched onto a bulk optics 50:50 beamsplitter and then imaged onto the CCD camera. The other arm propagates through single-mode fiber before being launched into free-space with another 10x microscope objective, and is superposed with the beam following the state generator at the 50:50 beamsplitter placed before the camera. By interfering a spherical wave of the fundamental Gaussian mode with a higher order mode generated by our platform, it is possible to visualize the interferogram and the intensity profiles on the camera \textbf{[21, 22, 30]}. A manual polarization controller and a fiber delay line is used in the upper arm to ensure indistinguishability of the two paths on the beamsplitter. A variable attenuator is also employed to ensure equal optical power in both arms before the beamsplitter.

With the setup described above we are able to measure the amplitude profile of each spatial mode (by blocking the upper path), or the phase profile when both paths are allowed to interfere at the beamsplitter. We measure the eigenstates of the three possible mutually unbiased bases (MUBs) in the two-dimensional Hilbert space formed by the superpositions of linearly polarized modes in a few-mode fiber. The three bases are $\{ |\text{LP}_\text{11a}\rangle, |\text{LP}_\text{11b}\rangle\}$, $|\{\text{LP}_{+}\rangle,|\text{LP}_{-}\rangle\}$ where $|\text{LP}_{\pm}\rangle = (1/\sqrt{2}) (|\text{LP}_\text{11a}\rangle \pm |\text{LP}_\text{11b}\rangle )$ and $\{|\text{OAM}_+\rangle, |\text{OAM}_-\rangle\}$, where $|\text{OAM}_{\pm}\rangle = (1/\sqrt{2})(|\text{LP}_\text{11a}\rangle +e^{\pm i\pi/2} |\text{LP}_\text{11b}\rangle)$. The theoretical amplitude and phase profiles are shown in Fig.\textbf{[2]} for each of the two states of each MUB, as well as the experimental results obtained with the setup of Fig.\textbf{[1]} b), showing these results match the theoretical predictions, demonstrating the ability of our spatial state generator to be able to faithfully reproduce any photonic state $\psi$ in two-dimensional space. In order to quantify the similarity of the simulated and measured amplitude profiles we measure the correlation between the 2D-Fourier spectrum of the amplitude profiles. From this data we calculate an average correlation of $94.2\pm3.7\%$. The deviation can be explained by the use of different exposure settings and nonideal alignment of the  CCD camera. The interferometer containing the phase modulator $\phi_x$ is partially insulated on an optical table, and is stable enough for the measurements displayed in Fig.\textbf{[2]} to be performed. For applications requiring the superpositions $|\text{LP}_{\pm}\rangle$ and $|\text{OAM}_{\pm}\rangle$ to be stable over longer time periods, active phase stabilization can be carried out with the injection of a reference laser at a different wavelength in the interferometer and implementing an active electronic control system.

\begin{figure}[ht]
\centering
    \includegraphics[trim={4.2cm 12cm 4cm 3cm},clip,scale=0.8]{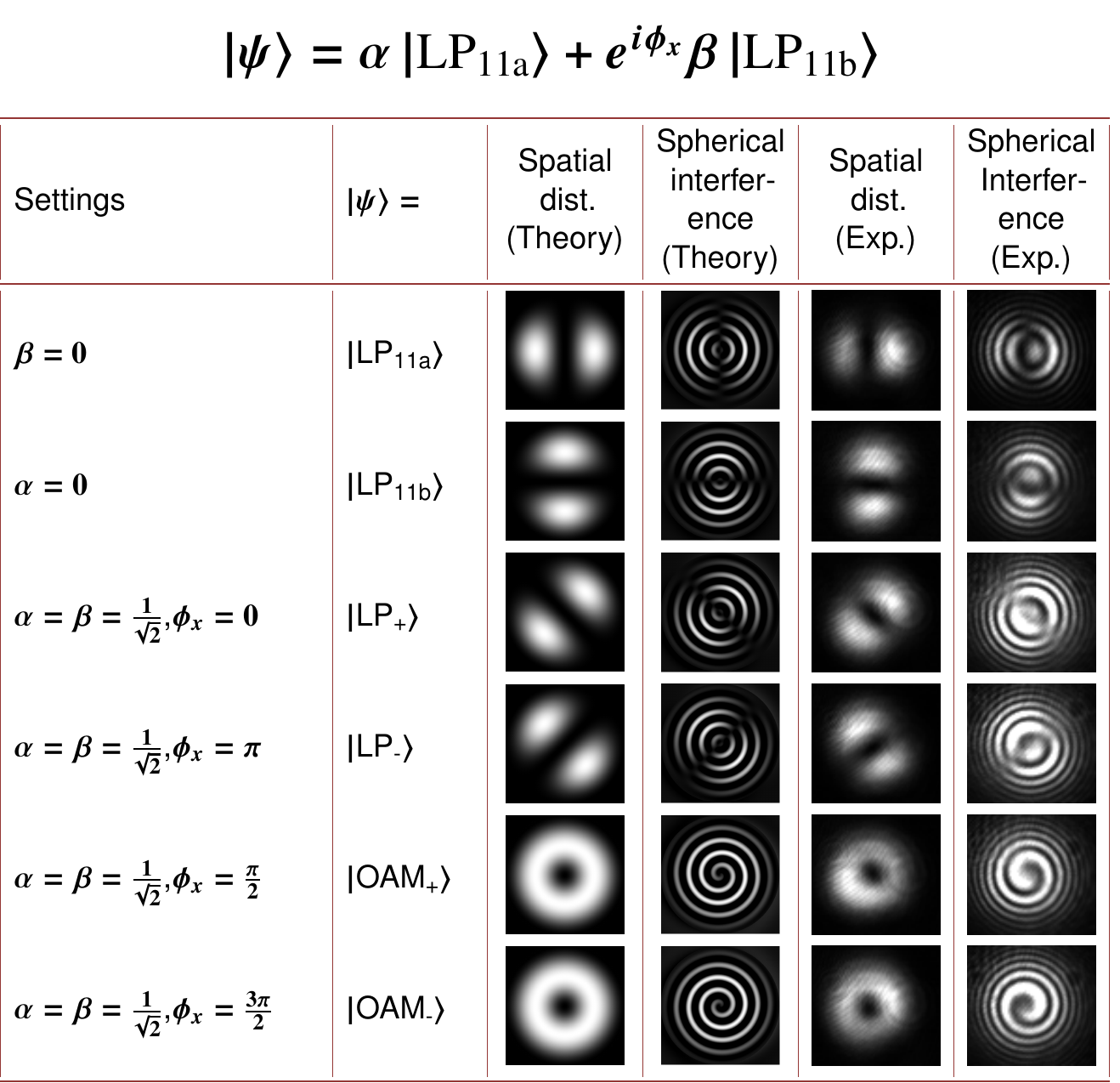}
    \caption{ 
    Theoretical and experimental amplitude and interferograms distributions of the different spatial states generated by our scheme, as imaged with an infrared InGaAs CCD camera (Fig. 1b))
 \label{fig:figure2}} 
\end{figure}

\begin{figure}[ht]
\centering
    \includegraphics[width=1\linewidth]{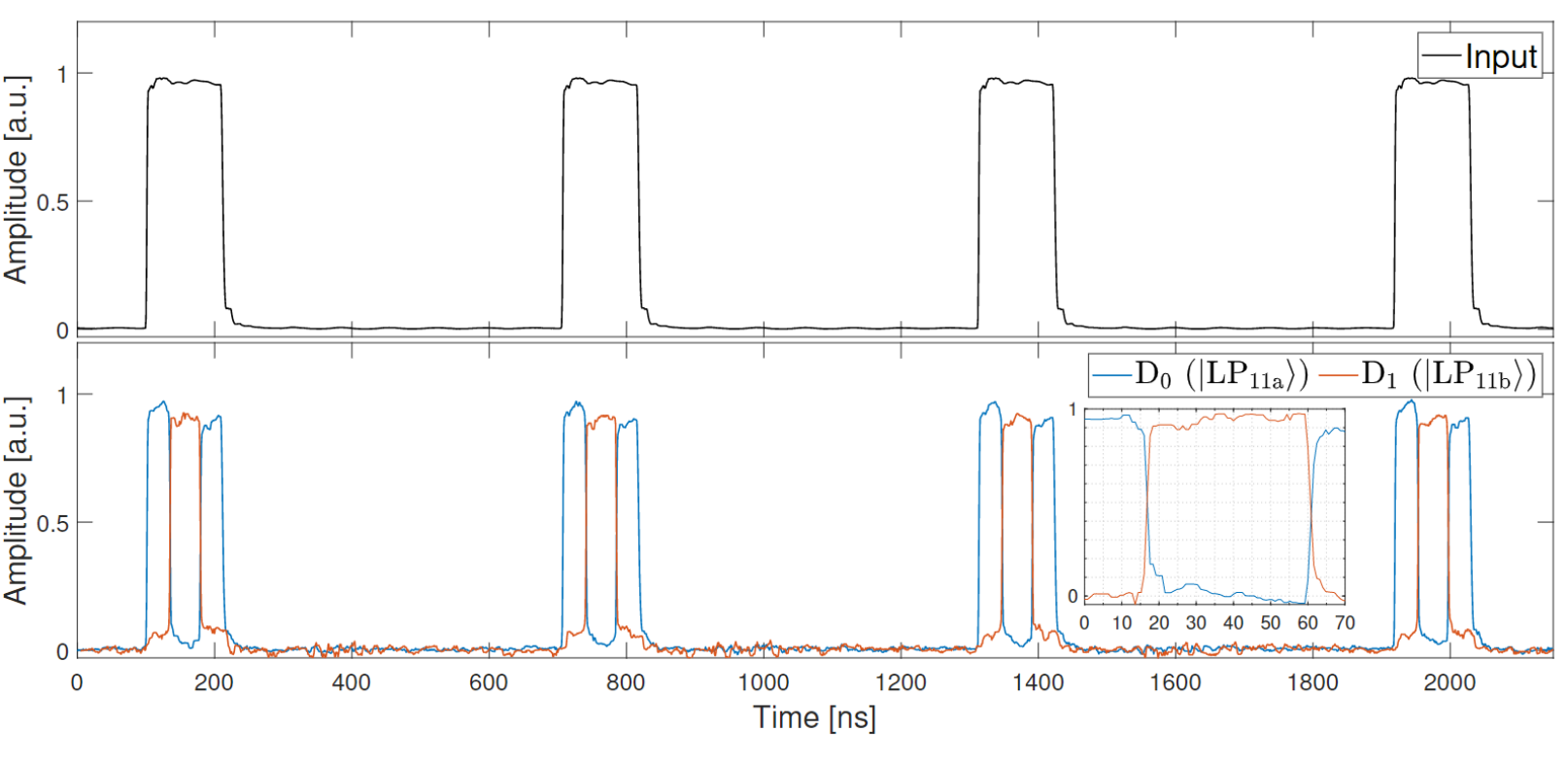}
    \caption{ Demonstration of dynamic switching. a) The 120 ns long optical pulses entering the Sagnac interferometer. b) Detected optical pulse after the second lantern  (PL2) when $\phi_R$ is modulated with a 45 ns wide pulse, centered within the 120 ns optical pulse c) Inset showing a high-resolution plot of the routing demonstrating a rise time and fall time of  $5.2$ ns and $2.4$ ns respectively.
 \label{fig:figure3}} 
\end{figure}


We now show the time response of the scheme when performing dynamic changes of photonic states (Fig. 3). We create 120 ns optical pulses which are sent to the Sagnac loop. We then generate a modulation pulse of 45 ns width by using a FPGA, which is used to drive $\phi_R$ inside the loop, adjusted to switch from $|\text{LP}_\text{11a}\rangle$ to $|\text{LP}_\text{11b}\rangle$ in the FMF. The time delay to the modulation of $\phi_R$ is synchronized with the 120 ns amplitude pulse such that only the middle portion of the optical pulse is routed in the Sagnac loop. We then measure the outputs of the second lantern (Fig. 1c)) with two amplified p-i-n photodiodes. We are thus able to show a switching of the spatial modes $|\text{LP}_\text{11a}\rangle$ to $|\text{LP}_\text{11b}\rangle$. The inset shows a zoom of the spatial state transition where a rise time of 5.2 ns and a fall time of 2.4 ns are shown. This response time is limited by our custom driving electronics and the bandwidth of our p-i-n diode.
However, our system can increase its performance even more using high-speed electronics as has been demonstrated in works using the Sagnac configuration \textbf{[31]}.

\newpage
\section{Demonstration of a spatial state quantum random number generator}

We then show that the dynamic and fast switching enabled by our spatial state generator allows the implementation of a measurement-device-independent (MDI) quantum random number generator (QRNG). QRNGs are essential to many applications, including simulations, online gambling and for cryptography \textbf{[32, 33]}. They are based on the fundamental principles of quantum mechanics, where the randomness does not depend on our own lack of knowledge regarding the physical process. One way to implement QRNG is through measurements on weak coherent states which is made by observing and recording the outputs on a measurement made on one of its degrees of freedom, for example, polarization \textbf{[34]}, the relative phase of two paths \\textbf{[35]} or arrival time at a detector \textbf{[36]}.

Traditional protocols for quantum random number generation depend on an implicit trust in the device used for state preparation and in the device used for measurement. While showing an advantage over deterministic methods for entropy generation and being comparatively practical to implement, they are prone to side-channel attacks, either in the detector or in the state preparation. To overcome the required trust in the devices, so called device-independent (DI) protocols have been developed which are able to generate genuine randomness while requiring no a priori assumptions about, or trust in, the devices. However, there are major technical challenges to overcome, and as such they are typically very slow compared to other implementations, and are not yet practical for most applications \textbf{[33]}. MDI protocols are a convenient alternative, as they are able to relax hardware requirements given the assumptions that Alice (the transmitter) has well characterized hardware responsible for state preparation while Bob (the measurement device) is untrusted \textbf{[27]}. Alice randomly chooses some experimental rounds to prepare and measure certain states that produce deterministic results at Bob in order to test the measurement device. It is then possible to bound the amount of information that has leaked to an eavesdropper, thus placing a lower bound on the amount of private bits that are being generated. In other rounds, in order to generate random numbers, Alice prepares a quantum state capable of generating a random output upon measurement projection. In this way, MDI protocols are highly applicable to practical situations where the measurement hardware cannot be fully trusted or characterized. 

Due to the dynamic nature of state preparation in MDI-QRNGs, our spatial state preparation platform is ideal to implement such a protocol. In our implementation the randomness comes from performing projective measurements of the spatial quantum superpositions of weak coherent states $|\psi\rangle$ onto the orthogonal elements $|\text{LP}_\text{11a}\rangle$ or $|\text{LP}_\text{11b}\rangle$ as Fig. Fig.\textbf{[1]} c) shows. The continuous wave laser is chopped up into 20 ns pulses at a repetition rate of $300$ kHz using an intensity modulator (omitted in Fig.\textbf{[1]} for brevity) and attenuated in order to create weak coherent states with an average photon number of 0.16 per detection gate at the output of the spatial state generator. Therefore, the contribution of multiphoton events is small in our experiment (< 1.15\%), and all double detection events are discarded. The output of our platform is connected to a second lantern PL2 in a back-to-back connection which makes the inverse mapping:  $|{\text{LP}_\text{11a}}\rangle \rightarrow |0 \rangle$ and $|\text{LP}_\text{11b}\rangle \rightarrow |1\rangle$, thus performing the measurement projection in the basis spanned by $|\text{LP}_\text{11a}\rangle$ and $|\text{LP}_\text{11b}\rangle$. Note that if we adjust in the spatial state generator $\alpha$ and $\beta$ to have the same value by adjusting the driving voltage of $\phi_R$ the incoming weak coherent state will have the same probability of being projected at port 1 or port 2 of PL2. The measurement performed in this fashion is equivalent to measuring in the path (computational) basis, and is thus independent of $\phi_x$. Thus the MDI-QRNG is independent of the phase instability of the interferometer at the input of the lantern. This driving voltage level is fine-tuned using the FPGA in such a way that the entropy of the generated random numbers is maximized. Following the same procedure, it is also possible to obtain either $\alpha=0$ or $\beta=0$ in order to carry out deterministic measurement outputs by setting $\phi_R$ accordingly, in which case the entropy of the generated bit sequence is minimized. We employ a single-detector scheme with time multiplexing in order to be able to measure both outputs \textbf{[35}, which implies in an extra 50\% loss at the detection stage. This means that the two outputs D$_0$ and D$_1$ in Fig.\textbf{[1]} c) are mapped to an early and a late time-bin which are then detected serially by one single-photon detector. We employ a single IdQuantique id210 single-photon detection module working in gated mode, with 10\% detection efficiency, a detection gate width of 5 ns, and dark count probability of 0.005$\%$. The total losses for the spatial state generator add up to approximately 14 dB, where 3 dB come from the two passes through the circulator, 3 dB for each phase modulator, 4 dB for the photonic lantern and an extra 1 dB for connector losses. We stress that these losses are not an issue for a weak coherent state source such as ours, since they are all concentrated on the transmitter. For the receiver an extra 4 dB loss comes from the second photonic lantern.

We run the MDI-QRNG protocol by randomly choosing in each measurement block whether to run test or randomness generation mode, with probabilities of 5\% and 95\% respectively. This random decision is given by a classical random number generator based on shift-registers implemented in the FPGA. We note that the additional source of randomness needed to run the MDI protocol is chosen by the user, based on something they can trust. Furthermore, under test mode, each of the two deterministic states is chosen with 50\% probability, such that the test states $|\text{LP}_\text{11a}\rangle$ or $|\text{LP}_\text{11b}\rangle$ are prepared with 2.5\% overall probability  on average. The test data is used to estimate the fraction of private bits generated by Alice and Bob, while the randomness mode data will generate the final output after randomness extraction. Depending on the measurement mode, different voltages are applied to the PM to modify $\phi_R$ inside the Sagnac loop. Furthermore, the FPGA registers the detection events from the single photon detector, and assigns a binary '0' ['1'] if the photon impinges on the detector at the time slots correponding to the different outputs of the lantern. Each measurement block consists of 32 kilobits, which are stored in a buffer inside the FPGA before they are sent over Ethernet to a personal computer for randomness extraction and assessment. We adjust the voltage driving the phase modulator $\phi_R$ to having close to equal probability of detecting a 0 or a 1, i.e. maximize the entropy of the generated sequence. The results of the 15 hour measurement run are displayed in Fig. 4a, where we plot the raw 8-bit Shannon entropy of each 32 kbit block as a function of time, giving an average of $7.15\pm0.12$  bits/Byte, thus demonstrating the stability of the system, with the deviation from the ideal case given mainly by the crosstalk of the photonic lantern. We also display the probabilities to detect the orthogonal test states $|\text{LP}_\text{11a}\rangle$ or $|\text{LP}_\text{11b}\rangle$ over the entire run, with the inset showing a narrower time range for these probabilities. The probabilities are defined as $P_{|\text{LP}_\text{11a}\rangle} = \frac{N_0}{N_0+N_1}$ and $P_{|\text{LP}_\text{11b}\rangle} = \frac{N_1}{N_0+N_1}$ where $N_0$ [$N_1$] are the number of detections in D$_0$ [D$_1$] for each measurement block when the state $\ket{\text{LP}_\text{11a}} [\ket{\text{LP}_\text{11b}}]$ is prepared and measured, i.e. when the QRNG is operating in test mode. Fig. 4b shows the normalized counts across the two detection time slots, as a function of a complete sweep of the driving voltage of phase $\phi_R$.

\begin{figure}[ht]
\centering
    \includegraphics[width=1\linewidth]{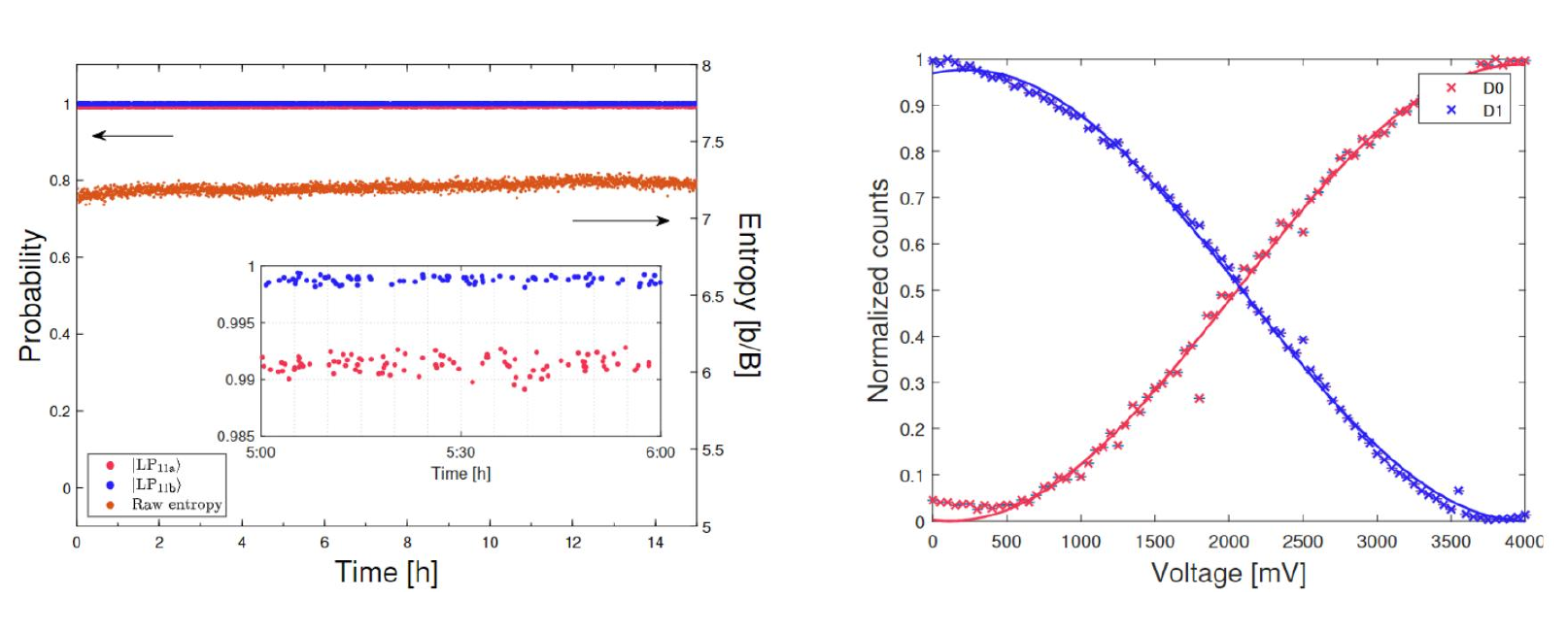}
    \caption{a) Probability of detecting a photon in D$_0$ [D$_1$] when the test states $\ket{\text{LP}_\text{11a}}$ [$\ket{\text{LP}_\text{11b}}$] are prepared and measured, and the raw entropy. Inset shows a zoom on the probabilities over 1 hour of the experimental run. b) Photon counts in D$_0$ [D$_1$] when sweeping the voltage applied to the PM inside the Sagnac loop. The non-ideal extinction of D$_0$ is a result of afterpulsing, which could be mitigated with longer time-bin separation
 \label{fig:figure44}} 
\end{figure}

 From the observed probabilities of recording a binary '0' ['1'] when working under test mode we are able to calculate an adversary's guessing probability $P_\text{guess}(\omega_x)$. In this case, $\omega_x$ can be any of the three states belonging to the set formed by $\{ |\text{LP}_\text{11a}\rangle, |\text{LP}_\text{11b}\rangle, |\phi\rangle \}$ where $|\phi\rangle$ is the superposition $|\phi\rangle = 1/\sqrt{2}(|\text{LP}_\text{11a}\rangle + e^{i\phi_x}|\text{LP}_\text{11b}\rangle)$, which is used as the randomness generation state. Regardless of the value of $\phi_x$ it will yield the same probabilities at the single-photon detectors, as the measurement is done in the path (computational) basis. Following the procedure in \textbf{[12]} we cast the problem into a simple convex optimization problem solved with semi-definite programming. We optimize over all POVMs that encode all possible guessing strategies an adversary Eve could use that are compatible with the observed probabilities in our experiment. From the observed success probability $P_{|\text{LP}_\text{11a}\rangle} = 0.9973 \pm 0.0144$ and $P_{|\text{LP}_\text{11b}\rangle}=0.9913\pm0.00093$ we compute Eve's guessing probability to be $0.64687$. Utilizing the fact that the number of private bits is given by $H_{\text{min}}(x^*) = -\log_2 P_\text{guess}(x^*)$ we are able to certify that at least $60.52$\% of the bits are private, by assuming all two-photon events are lowering the amount of generated privacy.
 
 Once the randomness extraction is done through a universal hashing extractor (Toeplitz extractor), following the methodology in \textbf{[35, 37]}, we run the generated sequence through the NIST 800-22 statistical test suite with the results displayed in Table 1. Each of the 15 NIST tests is run on 1 Mbit blocks, and as we can see, the average p-values for all blocks in each of the tests are clearly higher than the significance level of 0.01, meaning that our sequence is random within the confidence limits. Furthermore the proportion of passes for each 1 Mbit block within each test is higher than the confidence value of $0.98$, based on the number of trials that are done within each test.

\begin{table}

\begin{center}
\begin{tabular}{ l l l l }
 \textbf{Test} & \textbf{Mean p-value} & \textbf{Proportion pass} & \textbf{Verdict} \\ 
     \hline
Frequency & 0.51106 & 0.993 & PASS \\
BlockFrequency & 0.50021 & 0.995 & PASS \\
CumulativeSums & 0.50844 & 0.99 & PASS \\
Runs & 0.48528 & 0.991 & PASS \\
LongestRun & 0.49561 & 0.994 & PASS \\
Rank & 0.50007 & 0.991 & PASS \\
DFT & 0.49024 & 0.988 & PASS \\
NonOverlappingTemplate & 0.50032 & 0.987 & PASS \\
OverlappingTemplate & 0.49752 & 0.988 & PASS \\
Universal & 0.48368 & 0.991 & PASS \\
ApproximateEntropy & 0.50955 & 0.988 & PASS \\
RandomExcursions & 0.50661 & 0.98371 & PASS \\
RandomExcursionsVariant & 0.49858 & 0.99186 & PASS \\
Serial & 0.49774 & 0.985 & PASS \\
LinearComplexity & 0.50497 & 0.993 & PASS \\
\end{tabular}
\end{center}
\caption{Output of NIST test}

\end{table}
\bigskip
\section{Conclusion}
We have designed and tested an all-fiber platform that is able to generate any two-dimensional general transverse spatial state based on linear polarized modes in few-mode optical fibers. Our scheme is based on spatial division multiplexing technology, and is a good candidate as an ultra-fast generator of transverse spatial states, such as orbital angular momentum modes, for applications in classical and quantum communications. The speed of the system comes from the fact that it employs electro-optical telecom modulators, instead of bulk optical elements from other popular implementations. This capability distinguishes our approach from what has been done before, making this proposal attractive in any application where fast switching of spatial states is needed.  Furthermore, all the components used to develop the platform are commercially available, which is an advantage when replicating the results obtained in this work. 

We demonstrate the versatility of our platform in an implementation of an MDI-QRNG protocol, where dynamic switching of quantum states is needed. The system shows great stability over 15 hours of continuous operation. Our scheme can be further scaled to higher dimensions by using few-mode fibers \textbf{[13]}, beam splitters \\textbf{[12]}, and lanterns \textbf{[38]} that support a higher number of spatial modes. Although we have focused our efforts on showing that the platform has great potential to be employed as a basis for the integration of optical and quantum communications, many other areas can benefit directly from these results such as quantum imaging \textbf{[39]}, astronomy \textbf{[40]}, and metrology \textbf{[41]}.

\section{Acknowledge}
The authors acknowledge support from CENIIT Link\"{o}ping University, the Swedish Research Council (VR 2017-04470), the Knut and Alice Wallenberg Foundation through the Wallenberg Center for Quantum Technology (WACQT) and by the QuantERA grant SECRET (VR 2019-00392).

\end{document}